\documentclass[aps,pra,twocolumn,groupedaddress,showpacs,floatfix,superscriptaddress,longbibliography]{revtex4-2}
\usepackage[plainpages=false,pdfpagelabels,colorlinks=true,linkcolor=red,urlcolor=blue,citecolor=blue,pdftitle={Title},pdfauthor={},pdfdisplaydoctitle=true,pdfduplex=DuplexFlipLongEdge]{hyperref}
\usepackage{epsfig}
\usepackage{amsmath,amssymb}
\usepackage{physics}
\usepackage{graphicx}
\usepackage[dvipsnames,usenames]{color}
\usepackage[normalem]{ulem}
\usepackage{soul} 

\begin{document}

\title{Charge Singlets and Orbital Selective Charge Density Wave Transitions}
\author{Yuxi Zhang}
\affiliation{Department of Physics, University of California, Davis, CA 95616,USA}
\author{Chunhan Feng}
\affiliation{Department of Physics, University of California, Davis, CA 95616,USA}
\author{Rubem Mondaini}  
\affiliation{Beijing Computational Science Research Center, Beijing 100084, China}
\author{G.G. Batrouni}
\affiliation{Universit\'e C\^ote d'Azur, CNRS, Institut de Physique de Nice (INPHYNI), 06000 Nice, France}
\affiliation{Centre for Quantum Technologies, National University of
  Singapore, 2 Science Drive 3, 117542 Singapore}
\affiliation{Department of Physics, National University of Singapore, 2
  Science Drive 3, 117542 Singapore}
\affiliation{Beijing Computational Science Research Center, Beijing 100084, China}
\author{Richard T. Scalettar}
\affiliation{Department of Physics, University of California, Davis, CA 95616,USA}

\begin{abstract}
The possibility of `orbitally selective Mott transitions' within a multi-band
Hubbard model, in which
one orbital with large on-site {\it electron-electron repulsion} $U_1$ is insulating and another orbital, to which it is hybridized,
with small $U_{-1}$, is metallic, is a problem of long-standing debate 
and investigation.  In this paper we study an analogous phenomenon,
the co-existence of metallic and insulating bands in a system of orbitals with 
different {\it electron-phonon coupling}.
To this end, we examine two
 variants of the bilayer Holstein  model: a uniform bilayer and 
a `Holstein-metal interface' where
 the electron-phonon coupling (EPC), $\lambda$, is zero in the `metallic' layer. 
 In the  uniform bilayer Holstein model, charge density wave (CDW) order
 dominates at small interlayer hybridization  $t_3$, but decreases 
and eventually vanishes as $t_3$ grows, providing a charge analog of singlet
(spin liquid) physics. 
In the interface case, we show that CDW order penetrates into the 
metal layer and forms long-range CDW
 order at intermediate ratio of inter- to intra-layer hopping strengths,
$1.4 \lesssim t_3/t \lesssim 3.4$. 
This is consistent with the occurrence of an `orbitally selective CDW'
regime at weak $t_3$ in which the layer with $\lambda_{1} \neq 0$ exhibits 
long range charge order, but the `metallic layer' with $\lambda_{-1}=0$,
to which it is hybridized, does not.
\end{abstract} 

\maketitle

\section{Introduction}

Over the last several decades, much attention has been focused in the condensed matter community on 
layered materials.  One prominent example is that of the cuprate superconductors  
(SC)~\cite{bednorz1988,kastner1998,timusk1999,lee2006,vojta2009}.
   Bilayer graphene~\cite{dossantos2007,bistritzer2010,delaissardiere2010,xue2011,lang2012,mccann2013}
 is another, more recent, realization.
From a theoretical perspective, bilayer materials offer an opportunity to explore the
 competition between the formation of long range order at weak interlayer coupling and
 collections of independent local degrees of freedom in the limit of strong interlayer
 coupling. Computational studies have lent considerable insight into these phenomena, 
including quantitative values for the quantum critical
 points~\cite{sandvik1994,scalettar1994,vekic1995,wang2006,Golor2014} separating antiferromagnetic and singlet phases at zero temperature.
 
This competition is central to that which occurs in multiorbital systems,  notably the
 interplay of Ruderman-Kittel-Kasuya-Yosida order and singlet formation in the Kondo lattice
 and  periodic Anderson models~\cite{doniach1977,hess91,grewe91}. This close analogy originates
 in the observation that, in calculations on a model Hamiltonian, there is no difference 
between multi-layer and multi-orbital descriptions, apart from the interpretation of the
 additional label of the fermionic species. For this reason we will use the two terminologies
 interchangeably here. In multi-orbital language, one of the key conceptual interests 
is the possibility that the distinct values of the ratio of interaction strength to 
kinetic energy in the different bands might result in {\it separate} insulator 
transitions, i.e.~the possibility of an `orbital-selective' Mott transition (OSMT)
 ~\cite{liebsch2004,knecht2005,bouadim2009,koga2004,de2008,de2009,song2015,song2017,herbrych2019,herbrych2020,pandey2020,sroda2021}. 

Here we study analogous questions concerning bilayer (bi-orbital) systems in
 which the fermions interact with phonon degrees of freedom rather than via direct 
electron-electron correlations. A precise
 mathematical description of the mapping between the two situations is discussed. Quantum Monte Carlo
 (QMC) simulations have already been applied to the analysis of charge-density wave
 (CDW) and superconducting (SC) transitions in the single band Holstein
 model~\cite{holstein1959}. However, thus far, work has focused mostly on two-dimensional 
or three-dimensional models with a single kinetic energy 
scale~\cite{scalettar1989,marsiglio1990,weber2018,costa2018,hohenadler2019,li2019,zhang2019,chen2019,xiao2021,cohen2019,zhang2020,bradley2021,Johnston,feng2020i}. 

Using QMC simulations of the two-band Holstein model at half-filling, we will address
 the following questions concerning the effects of interband hybridization $t_3$: (i)  
Is there a transition in which CDW order is destroyed as $t_3$ is increased? What is 
the value of the critical coupling associated with the quantum critical point (QCP) in the ground state and the critical temperature for the thermal transitions at finite $T$? (ii) In a situation where
 the electron-phonon energy scales in the two bands are very different, can CDW order 
in one band coexist with metallic behavior in the other? These issues are in direct
analogy with those addressed in multiband Hubbard Hamiltonians; we will discuss 
similarities and differences between the resulting phenomena.

The paper is organized as follows: Sec.~\ref{sec:model} starts by describing the model Hamiltonian, with Sec.~\ref{sec:langevinalgorithm} introducing the numerical algorithm employed.
The main results of both the bilayer Holstein model and the Holstein-metal interface are given in Sec.~\ref{sec:holstein_bilayer} and Sec.~\ref{sec:holstein_interface},
 respectively. A contrast between the two models is drawn in Sec.~\ref{sec:spectral_nud}, by analyzing local quantities and gaps to excitations; Sec.~\ref{sec:summary}
 summarizes our results.

\section{Layered Holstein Hamiltonian}
\label{sec:model}

\begin{figure*}[t!]
\includegraphics[width=2\columnwidth]{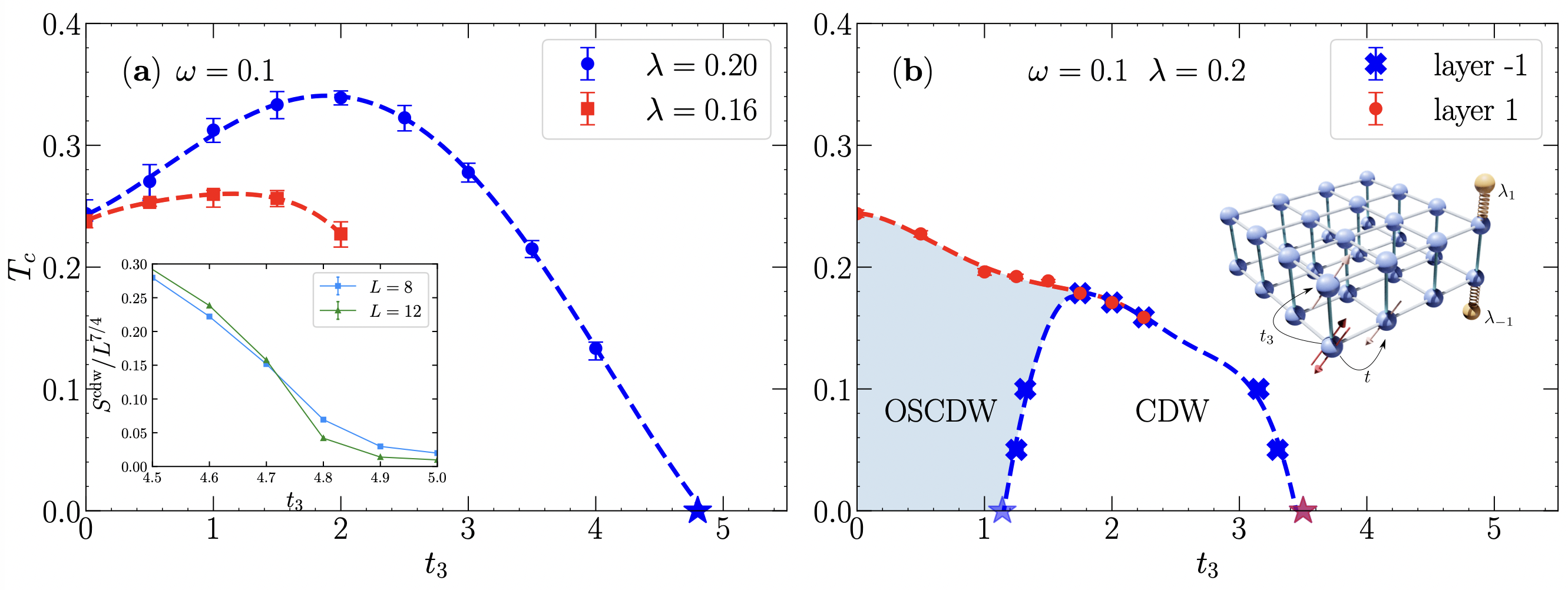} 
\caption{
(a)
Phase diagram of the Holstein bilayer giving the CDW transition temperature $T_c$ as a function of inter-layer
hopping $t_3$. 
Two values of electron-phonon coupling, 
$\lambda=0.2$ and  $\lambda=0.16$, are shown.
Data points are obtained by finite size scaling (FSS) analysis as shown in Fig.~\ref{fig:bilayerFSS}.
Dashed lines are guides to the eye.
Inset shows crossing plot of $S^{\rm cdw}/L^{7/4}$ versus
$t_3$ at $\lambda=0.2$ and low temperature $\beta=20$ and determines the
quantum critical value for the CDW-charge singlet transition.
(b)
Analog of (a), but for the Holstein-metal interface.
The four data points (blue crosses) at $T=0.1$ and $T=0.05$ are obtained by analysis
of the scaled structure factor, Fig.~\ref{fig:scalingQCP}.
Inset shows a sketch of a bilayer with relevant terms in Eq.~\eqref{eq:Holst_hamil} marked. 
QCPs are marked by stars on the $t_3$-axes in panels (a) and (b). 
The phonon frequency $\omega=0.1$ for all data.
}
\label{fig1} 
\end{figure*}

We focus on the bilayer Holstein model 
\begin{align} \label{eq:Holst_hamil}
\mathcal{\hat H} = & - \sum_{\langle ij
  \rangle,l, \sigma} \big(t^{\phantom{\dagger}}_{l}
\, \hat c^{\dagger}_{il \sigma} \hat
c^{\phantom{\dagger}}_{jl \sigma} + {\rm h.c.} \big) - 
\sum_{i,l, \sigma} \mu_{l} \hat n_{il \sigma}
\nonumber \\
& + \frac{1}{2 M} \sum_{il} \hat{P}^{2}_{il} +
\frac{1}{2} \sum_{i,l} \omega_{l}^{2}
\hat{X}^{2}_{il} + \sum_{i,l, \sigma} \lambda_{l} \, \hat
n_{il \sigma} \hat{X}_{il} 
\nonumber \\
& - \sum_{i,\langle l l'
  \rangle, \sigma} \big(t^{\phantom{\dagger}}_{ll'}
\, \hat c^{\dagger}_{il \sigma} \hat
c^{\phantom{\dagger}}_{il' \sigma} + {\rm h.c.} \big) \,\,.
\end{align}
$\hat c^{\phantom{\dagger}}_{il \sigma}(\hat c^{\dagger}_{il \sigma})$ are 
annihilation (creation) operators for an electron on layer $l(=\pm 1)$, 
site $i$ with spin $\sigma$, and $\hat n_{il \sigma}=
 \hat c^{\dagger}_{il \sigma}\hat c^{\phantom{\dagger}}_{il \sigma}$
 is the number operator. $t_l$ and $t_{ll'}=t_3$ denote the intra-
 and inter-layer hopping  respectively. Phonons are represented by 
local (dispersionless) quantum harmonic oscillators with frequency $\omega_{l}$,
 and on-site electron-phonon interaction on layer $l$ is introduced
 via $\lambda_{l}$. We choose intralayer hopping $t_l=t=1$ throughout this work to set
 the energy scale, and all simulations are done at half-filling
 $\langle \hat n_{il} \rangle=1$, which can be achieved
 by setting the chemical potential $\mu_{l}=-\lambda_l^2/\omega_l^2$; phonon mass is set as $M=1$. 
Each layer is an $L \times L$ site square lattice, as sketched in the inset of Fig.~\ref{fig1}(b), with $N=2 \times L \times L$ being the total number of sites.
We focus on two cases in this work: a uniform bilayer Holstein model where 
$t_1=t_{-1}=t$, $\mu_1=\mu_{-1}=\mu$, $\omega_1=\omega_{-1}=\omega$
 and $\lambda_{+1}=\lambda_{-1}=\lambda$; and an interface between Holstein layer and ``metal" layer, where only layer $l=+1$ has a non-zero electron-phonon coupling $\lambda_{+1} \neq 0$  and layer $l=-1$ has $\lambda_{-1}=0$.
We employ a recently developed Langevin quantum Monte Carlo (QMC) method~\cite{batrouni2019}
discussed in the next section.

We first define the local observables including the (layer-dependent) 
double occupancy,
\begin{align}
{\cal D}_l \equiv \langle \hat n_{il\uparrow} \hat n_{il\downarrow} \rangle 
\label{eq:D}
\end{align}
the near-neighbor intra-layer Green's function,
\begin{align}
{\cal G}_{\langle ij \rangle l} \equiv 
- \langle \hat c^{\dagger}_{il\sigma} \hat c^{\phantom{\dagger}}_{jl\sigma} 
+ \hat c^{\dagger}_{jl\sigma} \hat c^{\phantom{\dagger}}_{il\sigma}  \rangle  \, ,
\label{eq:Gintra}
\end{align}
and the near-neighbor inter-layer Green's function,
\begin{align}
{\cal G}_{\langle ll' \rangle} \equiv 
- \langle \hat c^{\dagger}_{il\sigma} \hat c^{\phantom{\dagger}}_{il'\sigma} 
+ \hat c^{\dagger}_{il'\sigma} \hat c^{\phantom{\dagger}}_{il\sigma}  
\rangle   \, .
\label{eq:Ginter}
\end{align}
When multiplied by their associated hopping integrals,
$t  \, {\cal G}_{\langle ij \rangle l}$ and
$t_3 \, {\cal G}_{\langle ll' \rangle}$
give the intra- and inter-layer kinetic energies per site.

Two further observables, the density-density and pair-pair
correlators, aid in characterizing the excitations between the planes. 
\begin{align}
d_{-1,1} & \equiv \frac{1}{4} \langle 
\hat n^{\phantom{\dagger}}_{i,1} \hat n^{\phantom{\dagger}}_{i,-1}-1 \rangle 
\nonumber \\
p_{-1,1} & \equiv -\frac{1}{4} \langle  
\hat \Delta^{\phantom{\dagger}}_{i,1}
\hat \Delta^{\dagger}_{i,-1} +
\hat \Delta^{\dagger}_{i,1} \hat \Delta^{\phantom{\dagger}}_{i,-1}  \rangle
\nonumber \\
\hat \Delta^{\dagger}_{il}  & \equiv
\hat c^{\dagger}_{il\uparrow}  \hat c^{\dagger}_{il\downarrow}.
\label{eq:singlet}
\end{align}
$d_{-1,1}$ and $p_{-1,1}$ are the analogs of the $zz$ and $xy$ spin correlations
which enter into the characterization of interlayer singlet formation in the 
Hubbard and Heisenberg bilayers, see Appendix \ref{sec:mag_lang}. Because of rotational symmetry of those models,
 their magnetic analogs, obtained by the transformation 
$\hat c^{\phantom{\dagger}}_{il\downarrow} \rightarrow \hat c^{\dagger}_{il\downarrow}$ 
are identical in value. $d_{-1,1} = p_{-1,1}$ would also hold in the attractive
 Hubbard Hamiltonian. Here, in the Holstein model, rotational symmetry 
is broken and we have $d_{-1,1} \neq p_{-1,1}$.  We will discuss the implications
 further in the sections to follow.

Characterization of the CDW formation in the thermodynamic limit can be made by the analysis of the 
(layer-resolved) structure factor, 
\begin{align}
S^{\rm cdw}_{l} = \frac{2}{N} \sum_{ij} (-1)^{i+j} \langle \hat n_{il} \hat n_{jl} \rangle,
\label{eq:Scdw}
\end{align}
with $\hat n_{il} = \sum_\sigma \hat n_{il\sigma}$. 
$S^{\rm cdw}_{l}$
samples correlations across the entire lattice, and hence is a primary tool in the determination of long range order.

In the case of the uniform bilayer, the quantities defined in Eqs.~\ref{eq:D}-\ref{eq:Ginter} 
and \ref{eq:Scdw} are independent of the layer index $l$, and in this
 case we suppress this index. But for the `interface' geometry,
 which includes one layer with $\lambda_1 \neq 0$ and another with $\lambda_{-1} = 0$,
 measurements performed on the two layers are inequivalent.

The layer-resolved single-particle spectral function $A_l(\omega)$ is obtained by using the maximum entropy method
to invert the integral equation relating the imaginary time dependent 
Green's function $G_{i=0}(\tau)$ and $A(\omega)$:
\begin{align}
G_{i=0}(\tau)&= \int d\omega  \frac{e^{-\tau \omega}} { 1 + e^{\beta \omega} }\, A(\omega)\nonumber 
\end{align}
\begin{align}
G_i(\tau)&= \langle \hat c^{\phantom{\dagger}}_i(\tau) 
 \hat c^{\dagger}_0(0)  \rangle
= \langle e^{\tau \mathcal H} \hat c^{\phantom{\dagger}}_i(0) e^{-\tau \mathcal H}  
 \hat c^{\dagger}_0(0)  \rangle \, .
\end{align}
$\tau$ represents imaginary time; layer and spin indices are omitted here for simplicity.
The appropriate local $G$ is used to get $A_l(\omega)$ for; each layer $l$ in the interface geometry.

We advance our key results in Figures \ref{fig1}(a) and \ref{fig1}(b):
 (i) At weak $t_3$ there is a phase transition at finite temperature $T_c$ to a state with long range
 charge order. In the bilayer case, $T_c$ initially increases with $t_3$ as the charge order 
 is enhanced by increased coordination number.
 (ii) At $T=0$, in both the Holstein bilayer
 and the Holstein-metal interface, CDW order is destroyed for $t_3$ exceeding a quantum critical value.
 (iii) The phase diagram of the interface geometry exhibits an `orbitally selective CDW phase' (OSCDW) at low $T$ and weak $t_3$. The specific description of how these phase diagrams are obtained is given in the corresponding section containing the main results of each model.
 
\section{Langevin Simulation Algorithm}\label{sec:langevinalgorithm} 

We employ a recently developed Fourier accelerated Langevin quantum Monte Carlo (QMC) method~\cite{batrouni2019}.
The partition function of the Holstein Hamiltonian is written as a path integral 
$\mathcal{Z} = \Tr e^{-\beta \mathcal{\hat H}} =
 \Tr e^{-\Delta \tau \mathcal{\hat H}}e^{-\Delta \tau \mathcal{\hat H} } \cdots e^{-\Delta \tau \mathcal{\hat H}}$
 where the inverse temperature $\beta=L_{\tau} \Delta \tau$ is discretized along
 the ``imaginary time'' axis. Complete sets of phonon eigenstates $|\{x_{i,\tau}\}\rangle$
 are inserted at each time slice, allowing the action of the phonon operators 
to be evaluated. In so doing, we convert the quantum problem into a classical problem 
in one higher dimension.  Since the fermion operators appear only as 
quadratic forms, we can trace over the associated degrees of freedom, leaving
 the partition function dependent only on the phonon field  $\{x_{i,\tau}\} $,
\begin{align}
\mathcal{Z} = \int \mathcal{D}x_{i,\tau} e^{-S_{\rm ph}} [\det M(\{x_{i,\tau} \})]^2
 = \int \mathcal{D}x_{i,\tau} e^{-S} \, ,
\end{align}
where the ``phonon action''
\begin{align}
S_{\rm ph}=\frac{\Delta \tau}{2} \left[\omega^2
\sum_{i}x_{i,\tau}^2 + \sum_{i}
\left(\frac{x_{i,\tau+1}-x_{i,\tau}}{\Delta \tau} \right)^2
\right] 
\label{eq:Sbose}
\end{align}
and
\begin{align}
   S = S_{\rm ph} - \ln (\det M)^2\,.
   \label{eq:S}
\end{align}
Here $M$ is a sparse matrix of dimension $NL_\tau$ whose detailed form is given in Ref.~\cite{batrouni2019}.
The square of the determinant appears because up and down fermionic species have the same coupling to
 the phonons. As a consequence, there is no sign problem~\cite{loh1990,mondaini2022}. In order to sample the phonon coordinates,
 instead of using the usual Metropolis algorithm, we evolve $ \{x_{i,\tau}\} $ using
 the discretized Langevin equation, 
 whose simplest form is given by the first order Euler discretization,
\begin{align}
    x_{i,\tau,t+dt} = x_{i,\tau,t} - dt \frac{\partial S}{\partial x_{i,\tau,t}}
 + \sqrt{2\,dt} \,\eta_{i,\tau,t}\,,
\end{align}
where 
{\it t} is the Langevin time, and $\eta$ is a Gaussian distributed
stochastic variable.
In practice, in our simulations we make use of a higher order Runge-Kutta discretization~\cite{batrouni2019} which reduces
the discretization error to ${\cal O}(dt^2)$. 
Throughout 
this work, the Langevin time step {\it dt} is chosen as 0.002, which has been shown to be sufficiently small so that the Langevin time discretization error 
is of the same order, or smaller than, statistical errors in typical simulations~\cite{batrouni2019}.
It can be demonstrated that, in the stationary limit, this Markov process generates configurations which are drawn from the exponential of the action of Eq.~\ref{eq:S}. The computational
 kernel is the calculation of the partial derivatives of the action via
\begin{align}
  \frac{\partial S}{\partial x_{i,\tau,t}} = \frac{\partial
 S_{\rm ph}}{\partial x_{i,\tau,t}} - 2 \Tr{\frac{\partial M}{\partial x_{i,\tau,t}}
 M^{-1}},
\end{align}
where the trace is evaluated using a stochastic estimator~\cite{batrouni2019}. 
Comparing to the conventional determinant quantum Monte Carlo (DQMC) method,
 which is an $O(N^3 L_\tau)$ approach, the Langevin method scales as $O(NL_\tau)$
 (although with a larger prefactor, so that there is a cross-over $N$ at which 
 the Langevin approach becomes the more efficient method). 
This enables simulations to reach considerably larger lattice sizes. 
In this work, we analyze systems up to $N=800$ sites. The efficiency 
of the Langevin approach results from the sparsity of the matrix $M$ 
and the fact that computing the action of $M^{-1}$ on a vector can be
 done iteratively in a number of steps which does not grow with
 $N$~\cite{batrouni2019}, with appropriate pre-conditioning. 
 The Langevin dynamics we employ is particularly effective in the adiabatic limit of small phonon-frequencies, where the
 density of zeros of individual fermion determinants is negligible~\cite{gotz2022}. In what follows we
  fix $\omega = 0.1$, thus simulations are stable, and statistical convergence is quickly obtained over
  the course of Markov generation.

\section{Holstein bilayer} \label{sec:holstein_bilayer}

We initially consider two identical layers with $\lambda=0.2, \, \omega=0.1$ and the question of the destruction of CDW order via the formation of charge singlets at large interlayer hopping $t_3$ before tackling the more complex issue of selective CDW transitions. 

\begin{figure}[t]
\includegraphics[width=1\columnwidth]{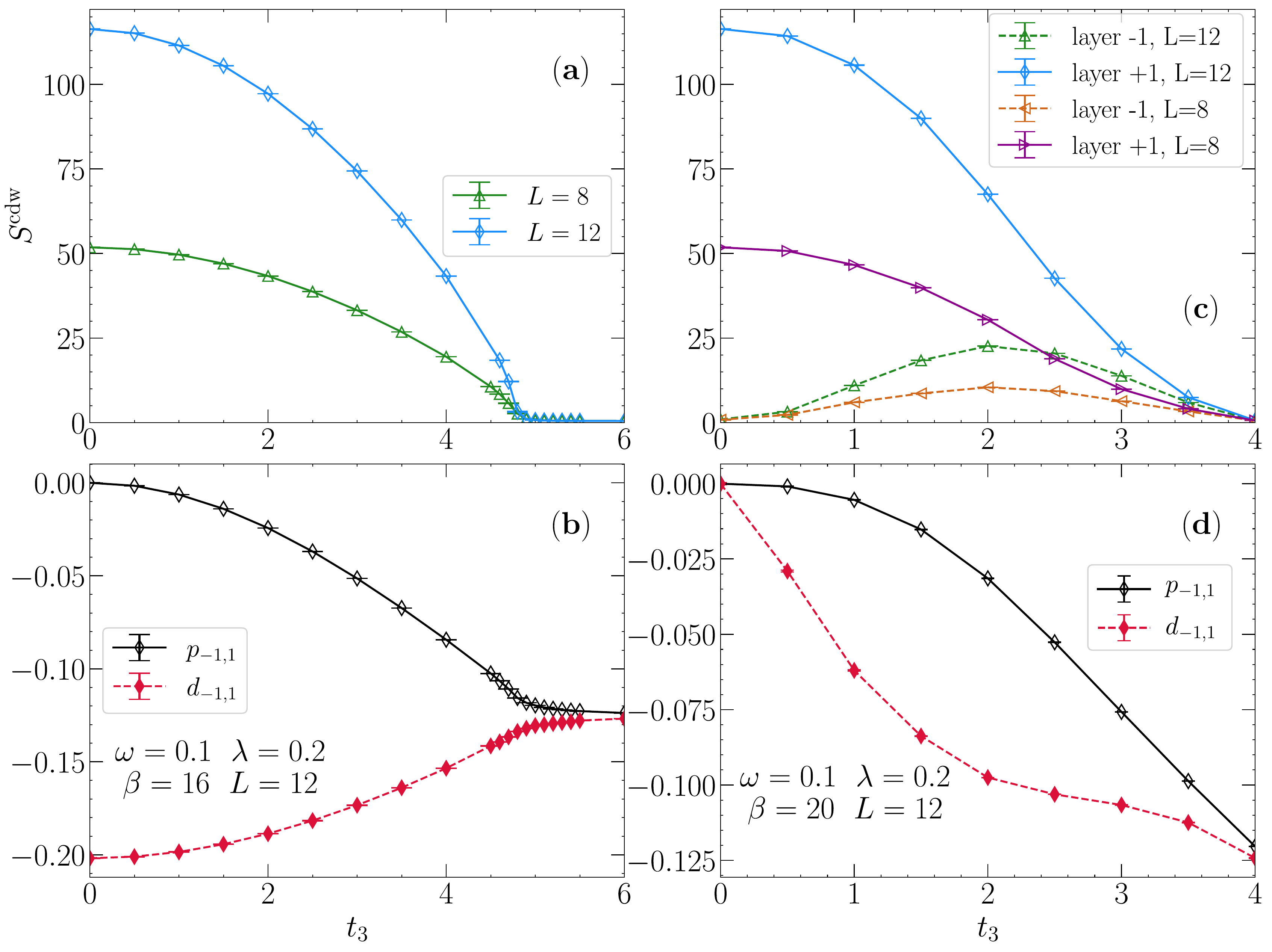} 
\caption{
(a) Charge structure factor $S^{\rm cdw}$; and (b) $p_{-1,1}$ and $d_{-1,1}$ as a function of interlayer hopping $t_3$ for the Holstein bilayer at low temperature $\beta=16$ and $\lambda_{+1}=\lambda_{-1}=0.2$.
$S^{\rm cdw}$ shows significant finite size effects in the ordered phase $t_3 \lesssim 4.8$.
Note that $d_{-1,1}$ vanishes at $t_3=0$, but jumps discontinuously to a non-zero value
for infinitesmal $t_3$.
Panels (c,d) are analog of (a,b) for the Holstein-metal interface.
The two curves in (c) correspond to layers $l=+1$ and $l=-1$, with
$\lambda_{+1}=0.2$ and $\lambda_{-1}=0$ respectively and temperature $\beta=20$.
In all plots the phonon frequency is set at $\omega=0.1$.
}
\label{fig:groundstate} 
\end{figure}

Figure \ref{fig:groundstate}(a) gives the CDW structure factor $S^{\rm cdw}$ as a function of $t_3$ at low temperature for two lattice sizes.  Below $t_{3,c} \approx 4.8$, $S^{\rm cdw}$ 
is large, and grows with lattice size, suggesting long range charge order. Figure \ref{fig:groundstate}(b) focuses on the interlayer density-density $d_{-1,1}$ and pair-pair $p_{-1,1}$ correlations. For small $t_3$, only $d_{-1,1}$ is large in magnitude, indicating coherence in the charge order between the two layers. As $t_3$ increases, intersheet pair correlations $p_{-1,1}$ develop. $d_{-1,1}$ and $p_{-1,1}$ then become nearly degenerate at $t_{3,c}$, signalling the loss of CDW order and entry into the `charge singlet' phase. Together, Figs.~\ref{fig:groundstate}(a) and \ref{fig:groundstate}(b) motivate the bilayer phase diagram of Fig.~\ref{fig1}(a). 

Although the QCP in this Holstein bilayer is closely analogous to that occurring in Hubbard and Heisenberg bilayers as well as the periodic Anderson model, in those cases the electron-electron interaction gives rise to magnetic phases which form due to the breaking of a {\it continuous} spin symmetry. Thus in 2D and quasi-2D geometries, no long range order is possible at finite $T$. In contrast, here for the Holstein model, charge and pairing order are not degenerate, as emphasized by the data of Fig.~\ref{fig:groundstate}(b). CDW correlations dominate at half-filling and a finite temperature phase transition can occur, terminating at a QCP as shown in Figs.~\ref{fig1}(a) and \ref{fig1}(b). This distinction means that, in principle, our characterization of the unordered phase as a `charge singlet' is somewhat loose: in the usual spin singlet the $x,y,z$ components of the spin-spin correlations on the two layers (or in the two orbitals) are equal. With that said, the equivalence of $p^{\phantom{\dagger}}_{-1,1}$ and $d^{\phantom{\dagger}}_{-1,1}$ in the large $t_3$ regime points to an emergent restoration of the symmetry (see Appendix \ref{sec:mag_lang}). It is worth noting that in the absence of $t_3$, e.g.~in the 2D Holstein model, this restoration does not occur until the anti-adiabatic limit is reached, which requires very large values of $\omega$~\cite{feng2020}.

\begin{figure}[t]
\includegraphics[width=1\columnwidth]{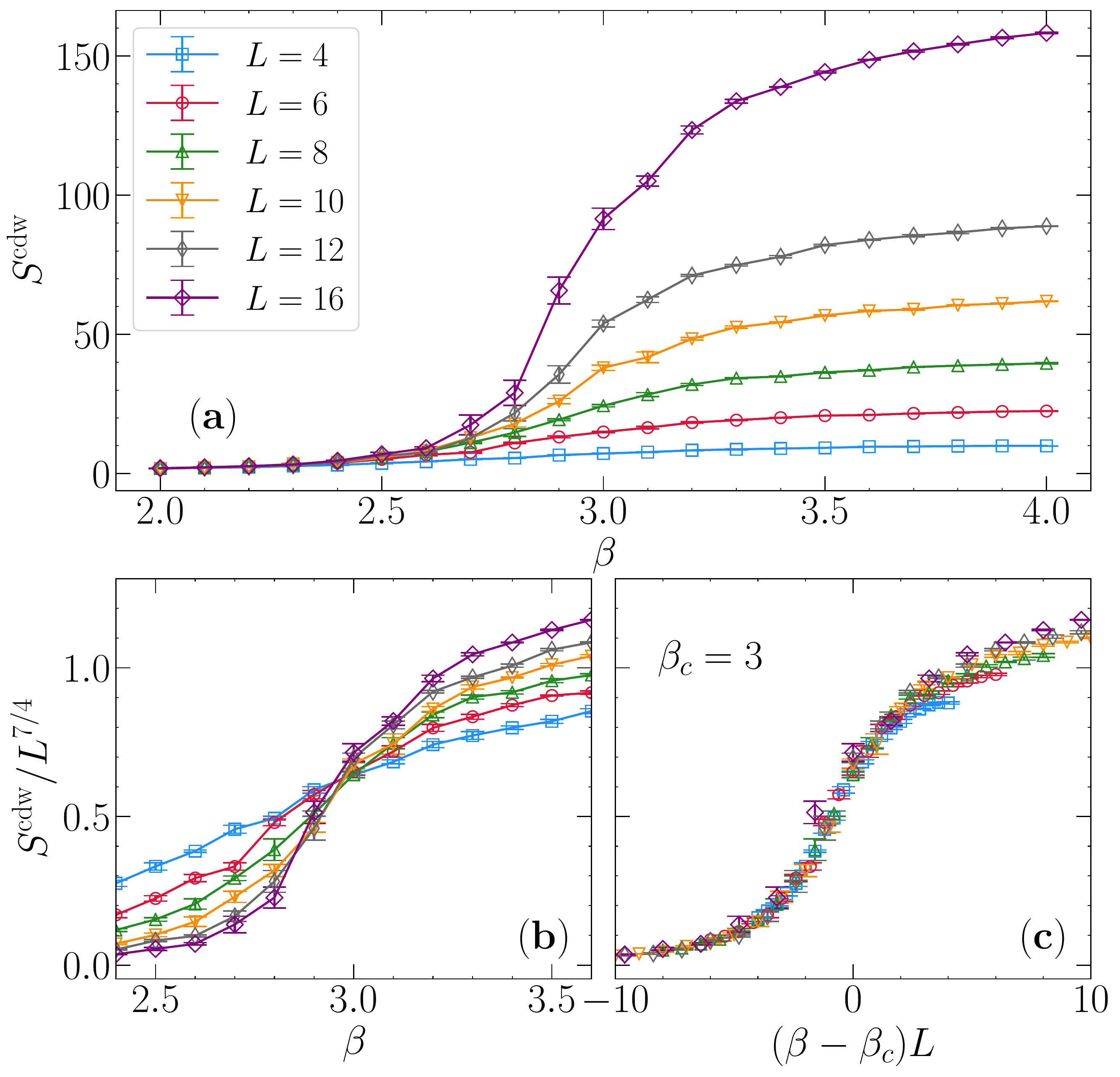} %
\caption{(a) CDW structure factor $S^{\rm cdw}$ dependence on the inverse 
temperature $\beta$ and finite size scaling of the Holstein bilayer at 
$t_3=2$. Both the crossing plot (b), and
the full data collapse (c) using 2D Ising critical exponents
and $\beta_c\simeq3.0$. 
 }
\label{fig:bilayerFSS} 
\end{figure}

Figure \ref{fig:bilayerFSS} provides details of the behavior of the CDW structure factor. The top panel (a) gives raw values for $S^{\rm cdw}$ as a function
 of $\beta$ at $t_3=2$ for different lattice sizes. At low $\beta$ (high temperature), the correlation length $\xi$  is short and $S^{\rm cdw}$ is independent
 of $L$. As $\beta$ increases, so does $\xi$ and when
$\xi \sim L$, $S^{\rm cdw}$ becomes sensitive to $L$. This separation of the curves provides a crude estimate for $\beta_c$, which may then be determined precisely by finite size scaling (FSS).

In particular, in the vicinity of the critical temperature $T_c$, the CDW structure factor measured on finite lattices of linear dimension $L$ should obey,
\begin{align}
    S^{\rm cdw} \sim L^{\gamma/\nu} f\left( \frac{T-T_c}{T_c} L^{1/\nu} \right)
    \,\,.
\label{eq:FSS}
\end{align}
As a consequence, when plotting $S^{\rm cdw}/L^{\gamma/\nu}$ as a function of the inverse temperature $\beta$, different sizes $L$ cross at $\beta=\beta_c$
 [Fig.~\ref{fig:bilayerFSS} (b)].  Following the scaling form given in Eq.~\eqref{eq:FSS} we note that when plotted against $(\beta-\beta_c)L^{1/\nu}$
 all data collapse on a single curve -- see  Figs.~\ref{fig:bilayerFSS}(c) (and later for the Holstein-metal interface in Fig.~\ref{fig:fulldatacollapse}).
 In this analysis we have used the critical exponents of the 2D Ising universality class ($\gamma=7/4$ and $\nu=1$), since the CDW phase breaks a $Z_2$
 symmetry. A discussion of the degree to which the collapse worsens, and hence the accuracy with which the exponents can be determined, is given in the
 Appendix~\ref{sec:crit_exp}.

Using such scaling procedure for various values of the interplane hybridization, allows us to extract the location of the thermal transition, as
 compiled in Fig.~\ref{fig1}(a), using two values of the electron-phonon interaction. In this geometry, the critical temperature $T_c$ initially
 increases as a consequence of the larger coordination number when the planes are coupled - the 2D to 3D crossover. However, at large $t_3$ the
 critical temperature decreases and ultimately vanishes at a quantum critical point. 


\section{Holstein-Metal interface} \label{sec:holstein_interface} 

We next consider the `Holstein-metal interface' in which layer $l=+1$ has nonzero $\lambda_{+1}$ but $\lambda_{-1}=0$. The two layers are in contact via
 hybridization.  Here, in addition to the question of charge singlet formation at large $t_3$, quenching CDW order, a different fundamental question arises:
 to what extent do CDW correlations in layer $l=+1$ `penetrate' into layer $l=-1$, and, conversely, is the CDW in layer $l=+1$ disrupted by contact with
 the `metallic' layer? We choose $\lambda=0.2$ and $\omega=0.1$ as in the previous section.

Figure \ref{fig:groundstate}(c) shows the CDW structure factor $S^{\rm cdw}$ in the two layers. $S^{\rm cdw}_{+1}$  decreases steadily with $t_3$: 
 additional quantum fluctuations associated with contact with the metal reduce charge order. In contrast, $S^{\rm cdw}_{-1}$ is non-monotonic: charge
 order is initially induced in the metal via contact with the Holstein layer, but ultimately large $t_3$ is inimical to it. The behavior of
 $S^{\rm cdw}_{-1}$  provides a first clue that order in layer $l=-1$ might occur only for intermediate $t_3$. Figure \ref{fig:groundstate}(d)
 gives the interlayer density-density $d_{-1,1}$ and pair-pair $p_{-1,1}$ correlations for this interface geometry. The primary difference
 from the original bilayer case is the gradual development of $d_{-1,1}$ with $t_3$. This is a consequence of the absence of CDW in the metal
 layer when $t_3$ vanishes. The interlayer hopping thus must not only couple the charge correlations, but also induce them in layer $l=-1$.
 Similar to the bilayer case, $d_{-1,1}$ and $p_{-1,1}$ become degenerate for large $t_3$. This is again a signature of entering into the charge singlet phase.

\begin{figure*}[t]
\includegraphics[width=2\columnwidth]{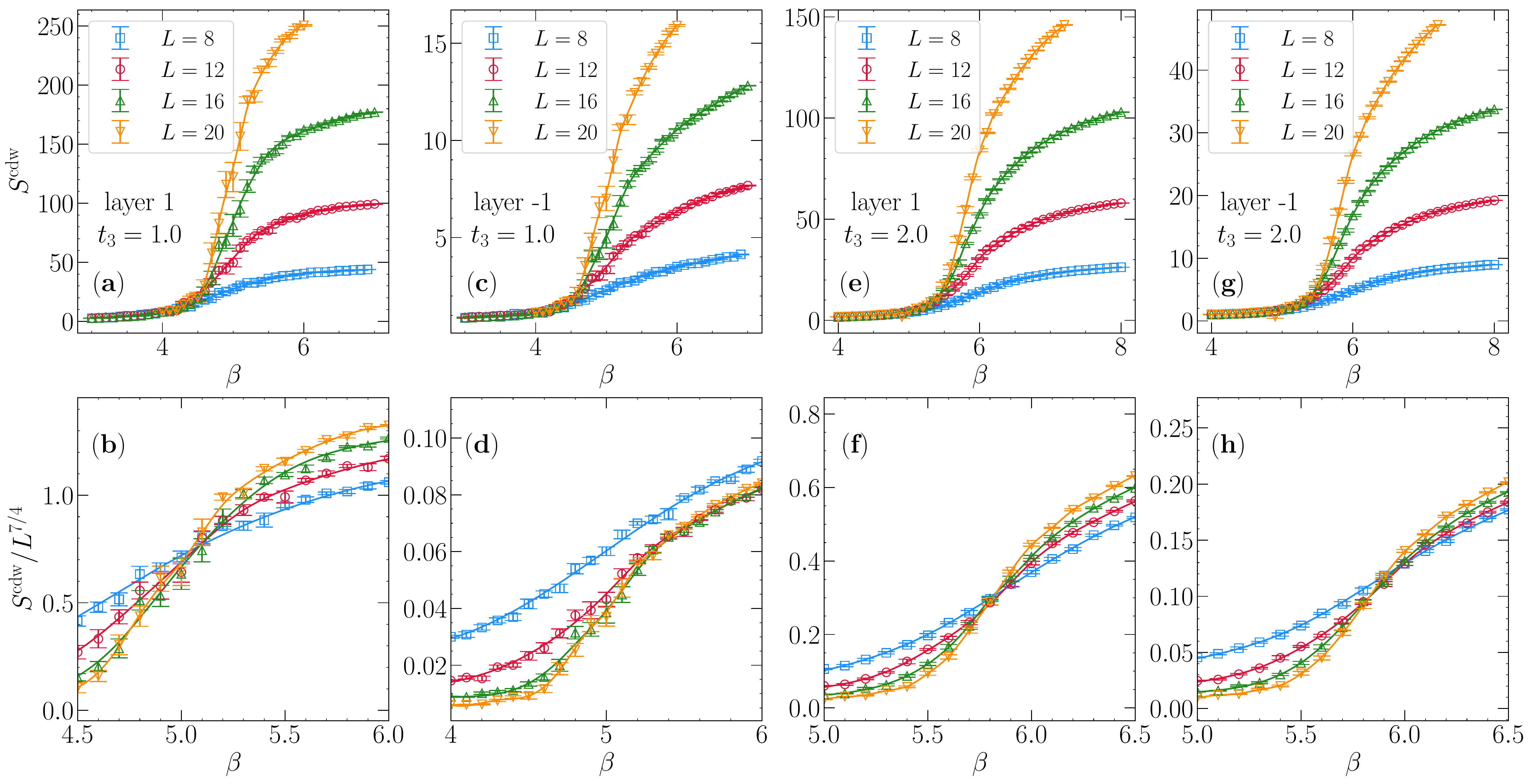}
\caption{(a) CDW structure factor, $S^{\rm cdw}$, dependence on the inverse temperature, $\beta$, for layer $l=1$ of the Holstein interface
 at $t_3=1$; (b) using 2D Ising critical exponents for finite size scaling (FSS). Panels (c) and (d) display the same but for the metallic
 layer $l=-1$. Panels (e--h) display the corresponding data for $t_3=2$. FSS in (f) and (h) show the same critical temperature for both
 layers at $t_3=2$, in contrast to $t_3=1$, where layer $l=+1$ (b) exhibits a clear CDW transition whereas data for layer $l=-1$ (d)
 does not exhibit crossing when using Ising critical exponents.
}
\label{fig:FSS} 
\end{figure*}

\begin{figure*}[t]
\includegraphics[width=2\columnwidth]{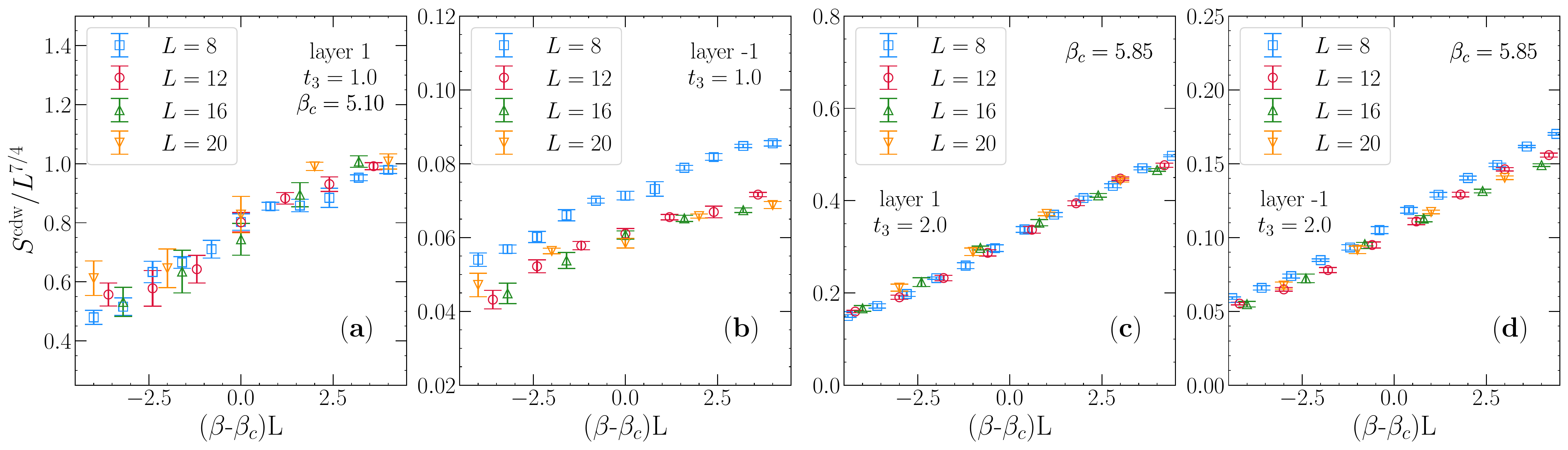} %
\caption{Full data collapse of the scaled CDW order parameter versus scaled reduced temperature in the `Holstein interface' system. Only the
 Holstein layer (layer 1) shows a single universal curve for $t_3=1$ (a), while both collapse for $t_3=2t$ (panels c,d).
 }
\label{fig:fulldatacollapse} 
\end{figure*}

We now turn to a more careful FSS study of the layer-resolved $S^{\rm cdw}_l$. Our main interest is in determining how long range order in the two layers
evolves with $t_3$. Figure \ref{fig:FSS} displays a detailed analysis of two representative values, $t_3=1$ and $t_3=2$. The former is a case when
 $S^{\rm cdw}_{-1}$ is just beginning to develop, and the latter is when $S^{\rm cdw}_{-1}$ has reached its maximal induced value [see
 Fig.~\ref{fig:groundstate}(c)]. There is a superficial resemblance in the unscaled data for both values of $t_3$, which rise as the temperature
 is lowered ($\beta$ increases) and also increase with system size. A proper scaling analysis, however, reveals a profound distinction. As seen
 in Figs.~\ref{fig:FSS}(b) and \ref{fig:FSS}(f), for both values of $t_3$, the layer $l=+1$ with non-zero electron-phonon coupling $\lambda_{+1}=0.2$,
 has a scaled structure factor  $L^{-7/4} S^{\rm cdw}_{+1}$ which exhibits a sharp crossing, indicating a finite temperature transition to long
 range CDW order.  When $t_3=2$, this crossing occurs  for the metallic layer with $\lambda_{-1}=0$ as well [Fig.~\ref{fig:FSS}(h)]. However,
 when $t_3=1$ the data for the metallic layer do not cross for the studied system sizes [Fig.~\ref{fig:FSS}(d)],
 namely $L=8\,$-$20$ for
 both $t_3$ values: The $L=12$,$16$ and $20$ curves converge at $\beta > 5.4$ instead of crossing. The $L=8$ data do not scale with the
 other lattice sizes at all.
 This distinction becomes even more apparent in Fig.~\ref{fig:fulldatacollapse}, where a simultaneous data
 collapse for the scaled structure factor can be made possible at the same temperature for $t_3=2$, while it is unattainable for $t_3=1$.

\begin{figure}[t]
\includegraphics[width=1\columnwidth]{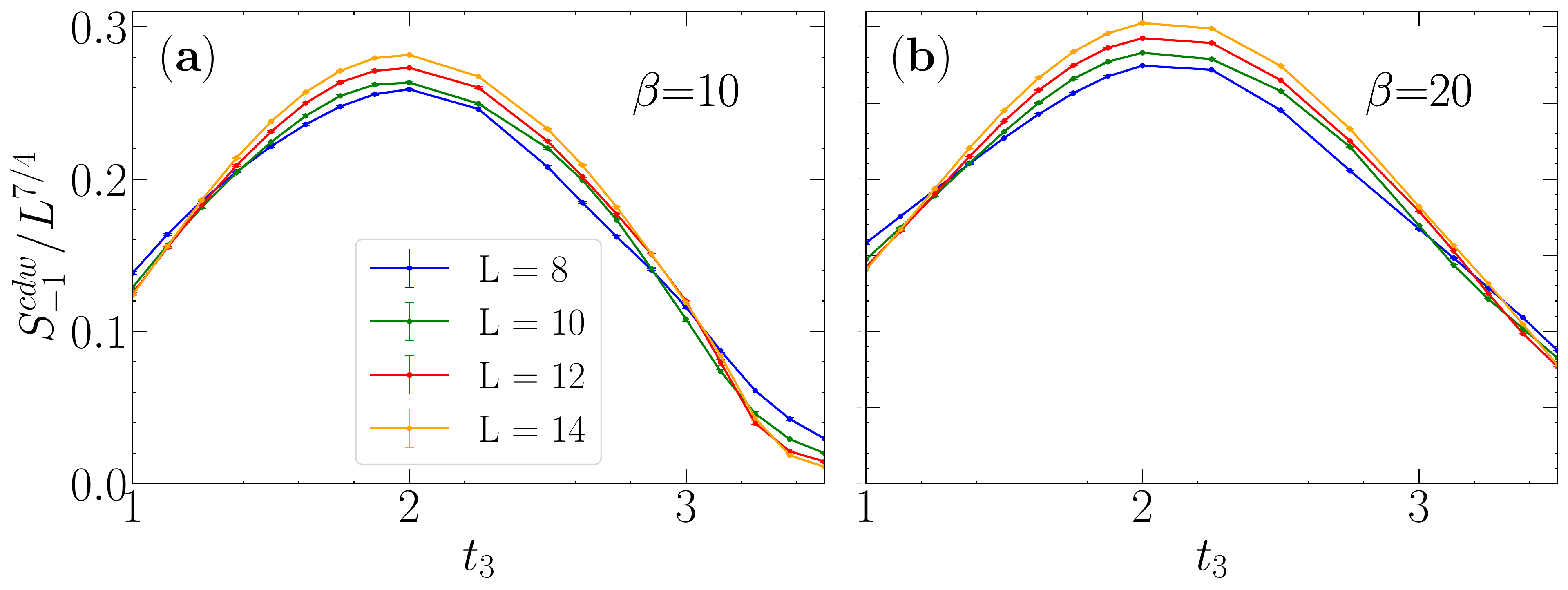} %
\caption{Crossing plot of $S^{\rm cdw}_{-1}/L^{7/4}$
(metallic layer) versus $t_3$ at $\lambda=0.2$ and $\omega=0.1$,
 at low temperature (a) $\beta=10$ (b) $\beta=20$, for the Holstein-metal
 interface. Two crossings are seen for in each case: (a) $t_3\sim 1.3$
 and $t_3\sim 3.2$ ($\beta=20$) and (b) $t_3\sim 1.33$ and
 $t_3\sim 3.1$ ($\beta=10$). 
All crossings are obtained with the critical exponents
 of the two-dimensional Ising model consistent with breaking
 the discrete
 $Z_2$ symmetry. The metallic layer, $l=-1$, is in the CDW phase
 only between the two values of $t_3$ where the curves cross.
 }
\label{fig:scalingQCP} 
\end{figure}

We conclude that for $t_3 = 2$, the interface geometry has CDW order in both
  layers, with long range correlations in the metallic layer induced
 by proximity to the Holstein layer. For $t_3 = t$, the interface geometry
  exhibits {\it orbital selective} CDW order- the metal remains  with
only short range correlation despite its hybridization to the long range CDW
  layer. This conclusion is supported by the
 data in Fig.~\ref{fig:scalingQCP}, where sweeps of the scaled $S^{\rm cdw}$
 with $t_3$ at two values $\beta=20,\,10$  show
 a pair of crossings. For $\beta=20$ these occur at
  $t_3 \sim 1.3$ and $t_3 \sim 3.2$ and for $\beta=10$ at $t_3 \sim 1.33$ and
  $t_3 \sim 3.1$. This analysis indicates that long range CDW order exists in
  layer $l=-1$ only between these values of $t_3$. 
  In particular, for $t_3$
  less than the lower critical value, layer $l=-1$ is not in a CDW phase
  while layer $l=+1$ is. 
  All crossings are obtained with critical exponents,
  of the two-dimensional Ising model, as 
  expected since a $Z_2$ symmetry is being broken.

\section{Spectral Functions and Double Occupancies}
\label{sec:spectral_nud}

\begin{figure}[b]
\includegraphics[width=1\columnwidth]{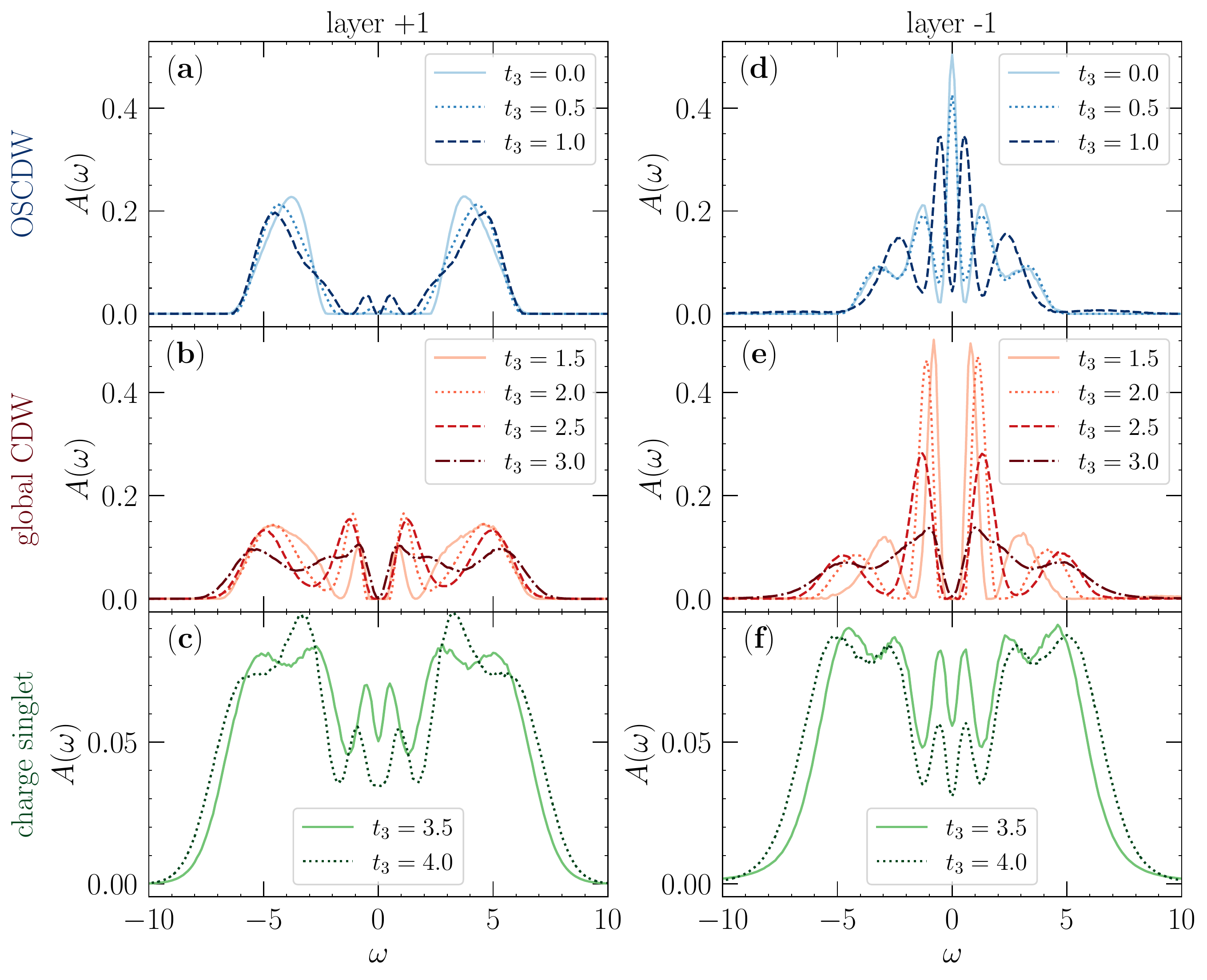} 
\caption{
Spectral function $A(\omega)$ at $\beta=12$ for several $t_3$ values cutting across
the Holstein-metal interface phase diagram of Fig.~\ref{fig1}.
\underbar{Top:}  Small $t_3$.
\underbar{Middle:}  Intermediate $t_3$.
\underbar{Bottom:}  Large $t_3$.
Left and right columns correspond to Holstein and metallic layers $l=+1$ and $l=-1$ respectively.
}
\label{fig:spectralfunction} 
\end{figure}

Having examined structure factors and inter-layer correlators, we now turn to the spectral functions and the double occupancies,
both of which provide additional insight into the ground state properties.
The layer-resolved spectral functions $A_l(\omega)$, shown in Fig.~\ref{fig:spectralfunction}, the many-body
 analog of the single particle density of states,
provide confirming evidence for the Holstein interface phase diagram of
Fig.~\ref{fig1}(b).
In the top row, for small $t_3$, 
the Holstein layer $l=+1$ exhibits a CDW gap. The gap at $t_3=0$ is large;
hybridization with the metal produces peaks closer to $\omega=0$, but a smaller gap remains.
On the other hand the metal layer $l=-1$ has finite Fermi surface spectral weight $A_{-1}(\omega=0) \neq 0$, thus showing the OSCDW.
In the middle row, for intermediate $t_3$, both layers have a gap, consistent with the measurement of
simultaneous long range CDW order.
Finally, in the bottom row, for large $t_3$, 
both layers have finite Fermi surface spectral weight $A_l(\omega=0) \neq 0$ for $l=+1,-1$.
The system is in the charge singlet (charge liquid) phase.

We note that although the bilayer and interface geometries have many
properties in common at large $t_3$, their spectral functions are different.
There is a gap in the bilayer case, but not for the interface.
We have verified, with separate exact diagonalization calculations, that for dimers (i.e.~$t_3 \gg t$) with
$\lambda_{+1}=\lambda_{-1}$ one finds $A(\omega)$ is gapped, while
when
$\lambda_{+1}$ is nonzero $\lambda_{-1}=0$ one reproduces the behavior shown in Fig.~\ref{fig:spectralfunction}(c,f).
\color{black}

\begin{figure}[t!]
\includegraphics[width=1\columnwidth]{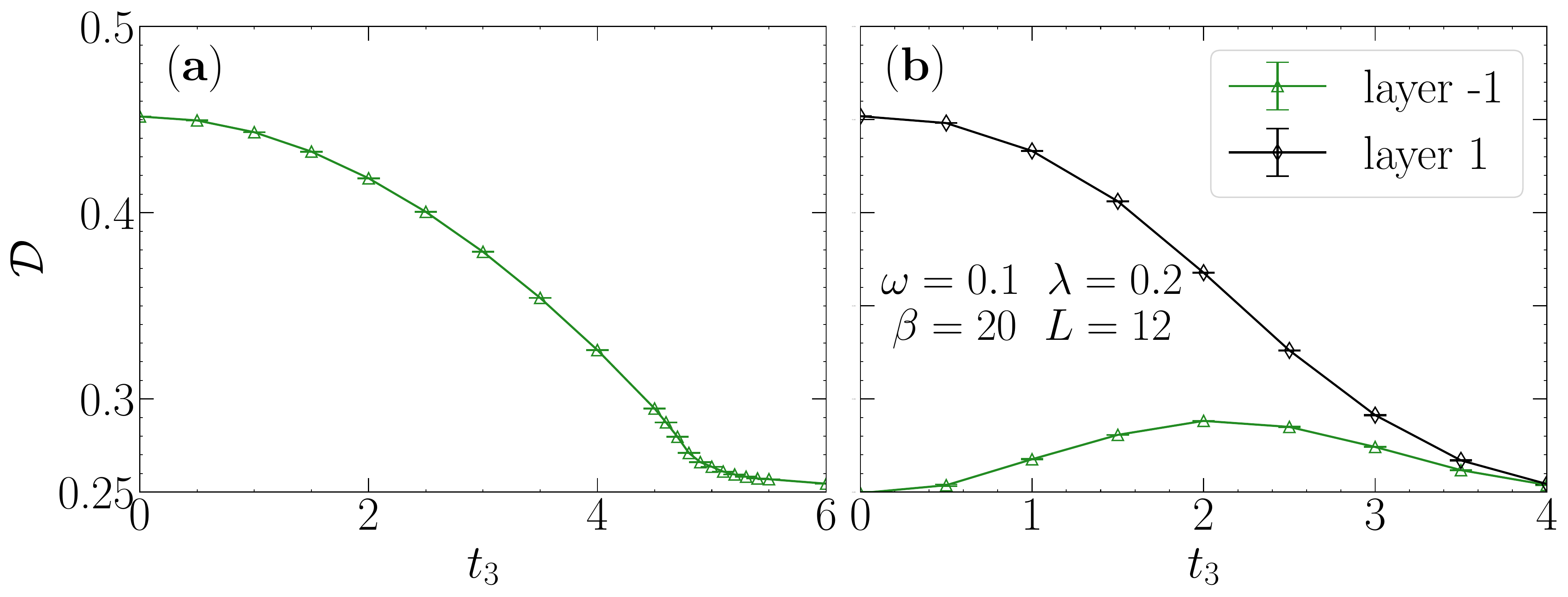} %
\caption{
Double occupancy ${\cal D}$
 shown as a function of interlayer hopping $t_3$ at low temperature $\beta=20$.
  ${\cal D}$ is a local observable, and its value is the same for $L=8$ and $L=12$ to 
  within the symbol size; we show the latter.
  (a)  Holstein bilayer;
  (b) Holstein-metal interface.
  In (a) the two layers are equivalent and a single curve is shown.  In (b) the
    green curve shows data on layer $l= -1$, whereas the grey curve represents
   layer $l=+ 1$.
 }
\label{fig:doubleocc} 
\end{figure}

In a perfect CDW phase, half of the sites are doubly occupied and half are empty,
and ${\cal D} = 0.5$. In the absence of interactions, $\lambda=0$, all four site occupation possibilities 
$|0\rangle, \,
 |\uparrow \rangle, \,
 |\downarrow \rangle,$ and
 $|\uparrow \downarrow \rangle$,
are equally likely and ${\cal D} = 0.25$. This is also the case in the charge singlet phase.
Figure ~\ref{fig:doubleocc} shows ${\cal D}$ as
  a function of $t_3$.  Panel (a) is for the bilayer, where 
 ${\cal D}_{+1}
 ={\cal D}_{-1}$, and panel (b) for the interface geometry where the two are inequivalent.
  In both cases,  ${\cal D}_{+1}$  is seen to evolve between the CDW and singlet limits, although it never attains the value ${\cal D}=0.5$ owing to the presence of quantum fluctuations.  For the interface, ${\cal D}_{-1}$ begins at the uncorrelated value, 
 ${\cal D}_{-1}=0.25$ at $t_3=0$ since $\lambda_{-1}=0$. The double occupancy evolution is quite similar to that of the structure factor, Fig.~\ref{fig:groundstate}(a,c). However,  since ${\cal D}$ is a local observable, it exhibits less sharp features than $S^{\rm cdw}$ 
 in the vicinity of the QCP and thus only provides qualitative evidence for a cross-over between those two phases. Besides that, a simple model which exhibits a layer-dependent trivial CDW formation with only electronic degrees of freedom, a bilayer ionic model, already displays this characteristic non-monotonicity with growing hybridization (see Appendix \ref{sec:analog_model}).

\section{Discussion and conclusions}
\label{sec:summary}

We have generalized our existing understanding of the effect of
 interlayer/interorbital hybridization $t_3$ on {\it magnetic} order 
driven by an on-site {\it electron-electron repulsion} in the Hubbard
 model to {\it charge} order originating in {\it electron-phonon interactions} 
in the Holstein model.  The two scenarios, although qualitatively 
related, are quite distinct in detail owing to the lower symmetry
 of the CDW order parameter relative to the magnetic case. 
Despite this difference, and its consequences such as the 
appearance of charge order at finite temperature, the basic
 feature of the destruction of long range order in the limit of large hybridization is shown still to occur.
Indeed, one remarkable conclusion of our work is that $t_3$
seems to restore the degeneracy of pairing and charge correlations at the QCP. 

Our most interesting observation is that the coexistence of CDW order on
a layer with non-zero electron-phonon coupling $\lambda$ with a metallic phase 
on the $\lambda=0$ layer, which is trivially true at $t_3=0$, likely extends
out to finite $t_3$.  This conclusion is based on the inability to scale
the charge correlations in the $\lambda=0$ layer unless $t_3 \gtrsim 1.4$ (for $\lambda_{+1}=0.2$).

The possibility that charge order takes place selectively parallels the known occurrence of distinct
Mott transitions in multi-orbital Hubbard models in the coexistence of
metallic and insulating behavior.  The connection is, however, not exact, since in principle a Mott transition
might occur in the absence of spontaneous symmetry breaking, whereas the insulating CDW phase here breaks $Z_2$ symmetry.
With that said, the Mott transition in its most common incarnation, the square lattice Hubbard model, {\it is}
always accompanied by long range antiferromagnetic order.
Thus our work does provide a close analog of the case of orbital selective transitions
in bands with differing electron-electron interaction strengths.

\begin{acknowledgments}
\noindent
The work of Y.X.Z., C.H.F., and R.T.S. was supported by the grant DE‐SC0014671 funded 
by the U.S. Department of Energy, Office of Science. R.M. acknowledges support from the NSFC Grants No. U1930402, No. 12050410263, No. 12111530010, No. 11974039, and No. 12222401.
\end{acknowledgments}

\appendix 

\section{Connection to Magnetic Language}\label{sec:mag_lang}

In the repulsive 2D Hubbard model the dominant physics at half-filling on a bipartite lattice
is anti-ferromagnetic order, characterized by the operators,
\begin{align}
\hat S^j_x &= \frac{1}{2} \big( \, \hat S^j_+ + \hat S^j_- \, \big)
= \frac{1}{2} \big( \,
\hat c^{\dagger}_{j\uparrow} \hat c^{\phantom{\dagger}}_{j\downarrow}
+ \hat c^{\dagger}_{j\downarrow} \hat c^{\phantom{\dagger}}_{j\uparrow} \, \big)
\nonumber  \\
\hat S^j_y &= \frac{1}{2i} \big( \, \hat S^j_+ - \hat S^j_- \, \big)
= \frac{1}{2i} \big( \,
\hat c^{\dagger}_{j\uparrow} \hat c^{\phantom{\dagger}}_{j\downarrow}
- \hat c^{\dagger}_{j\downarrow} \hat c^{\phantom{\dagger}}_{j\uparrow} \, \big)
\nonumber  \\
\hat S^j_z &=  
\frac{1}{2} \big( \, \hat n_{j\uparrow} - \hat n_{j\downarrow} \, \big) =\frac{1}{2}
\big( \, \hat c^{\dagger}_{j\uparrow} \hat c^{\phantom{\dagger}}_{j\uparrow}
- \hat c^{\dagger}_{j\downarrow} \hat c^{\phantom{\dagger}}_{j\downarrow} \, \big)
\label{eq:spinops}
\end{align}
From these relations, and as a consequence of the spin SU(2) symmetry of the Hubbard model,
\begin{align}
\langle \hat S^j_+ \hat S^i_- + \hat S^j_- \hat S^i_+  \rangle &=
4\, \langle \hat S^j_z \hat S^i_z \rangle 
\nonumber \\
\langle \, \hat c^{\dagger}_{j\uparrow} \hat c^{\phantom{\dagger}}_{j\downarrow} 
 \hat c^{\dagger}_{i\downarrow} \hat c^{\phantom{\dagger}}_{i\uparrow} 
+ \hat c^{\dagger}_{j\downarrow} \hat c^{\phantom{\dagger}}_{j\uparrow} 
\hat c^{\dagger}_{i\uparrow} \hat c^{\phantom{\dagger}}_{i\downarrow}  \, \rangle &=
\nonumber \\
 \langle \,
\big( \, \hat c^{\dagger}_{j\uparrow} \hat c^{\phantom{\dagger}}_{j\uparrow}
- \hat c^{\dagger}_{j\downarrow} \hat c^{\phantom{\dagger}}_{j\downarrow} \, \big)
&\big( \, \hat c^{\dagger}_{i\uparrow} \hat c^{\phantom{\dagger}}_{i\uparrow}
- \hat c^{\dagger}_{i\downarrow} \hat c^{\phantom{\dagger}}_{i\downarrow} \, \big)
\, \rangle 
\label{eq:spinsym}
\end{align}

If we perform a particle-hole transformation 
to the down spin fermions,
\begin{align}
\hat c^{\dagger}_{j\downarrow} &\,\, \rightarrow \,\,
(-1)^j \hat c^{\phantom{\dagger}}_{j\downarrow} 
\nonumber \\
\hat n_{j\downarrow} \,\, &\rightarrow \,\,
\big( \, 1 - \hat n_{j\downarrow} \, \big)
\nonumber \\
 \hat S^j_+ = \hat c^{\dagger}_{j\uparrow} \hat c^{\phantom{\dagger}}_{j\downarrow}
 &\,\, \rightarrow \,\, (-1)^j
 \hat c^{\dagger}_{j\uparrow} \hat c^{\dagger}_{j\downarrow}
 \equiv (-1)^j \hat \Delta^{\dagger}_j
\nonumber \\
\hat S^j_z =    \frac{1}{2} \big( \, \hat n_{j\uparrow} - \hat n_{j\downarrow} \, \big) 
&\,\, \rightarrow\,\,  
 \frac{1}{2} \big( \, \hat n_{j\uparrow} + \hat n_{j\downarrow} \, \big) 
 \equiv \hat n_j
\label{eq:pht}
\end{align}
we conclude that,
\begin{align}
 - \, \langle \hat \Delta^{\dagger}_j  \hat \Delta^{\phantom{\dagger}}_i 
 +  \hat \Delta^{\phantom{\dagger}}_i  \hat \Delta^{\dagger}_j  \rangle
 = \langle \big( \, \hat n_j - 1) \, \big) 
 \big( \, \hat n_i - 1) \, \big)  \rangle
\label{eq:chpairsym1}
\end{align}
assuming that sites $i$ and $j$ are on opposite sublattices.

If, finally, assuming we are at half-filling, so that $\langle \hat n_j \rangle = 1$,
\begin{align}
 - \, \langle \hat \Delta^{\dagger}_j  \hat \Delta^{\phantom{\dagger}}_i 
 +  \hat \Delta^{\phantom{\dagger}}_i  \hat \Delta^{\dagger}_j  \rangle
 = \langle \hat n_j \hat n_i  - 1 \rangle
\label{eq:chpairsym2}
\end{align}
This shows that the two correlation functions of 
Eq.~\ref{eq:singlet} are equal: $p_{1,-1} = d_{1,-1}$.
The merging of the two curves of 
Fig.~\ref{fig:groundstate}(c,d)
at a common value reflects a restoration of an SU(2) symmetry
of the Hubbard model.
It is interesting that this occurs even though the correlators are not in the singlet
 limit of $-1/4$ (due to the fact that we are not in Holstein analog of the large $U$ limit).

\section{Extracting the critical exponents} \label{sec:crit_exp}

\begin{figure}[t!]
\includegraphics[width=0.99\columnwidth]{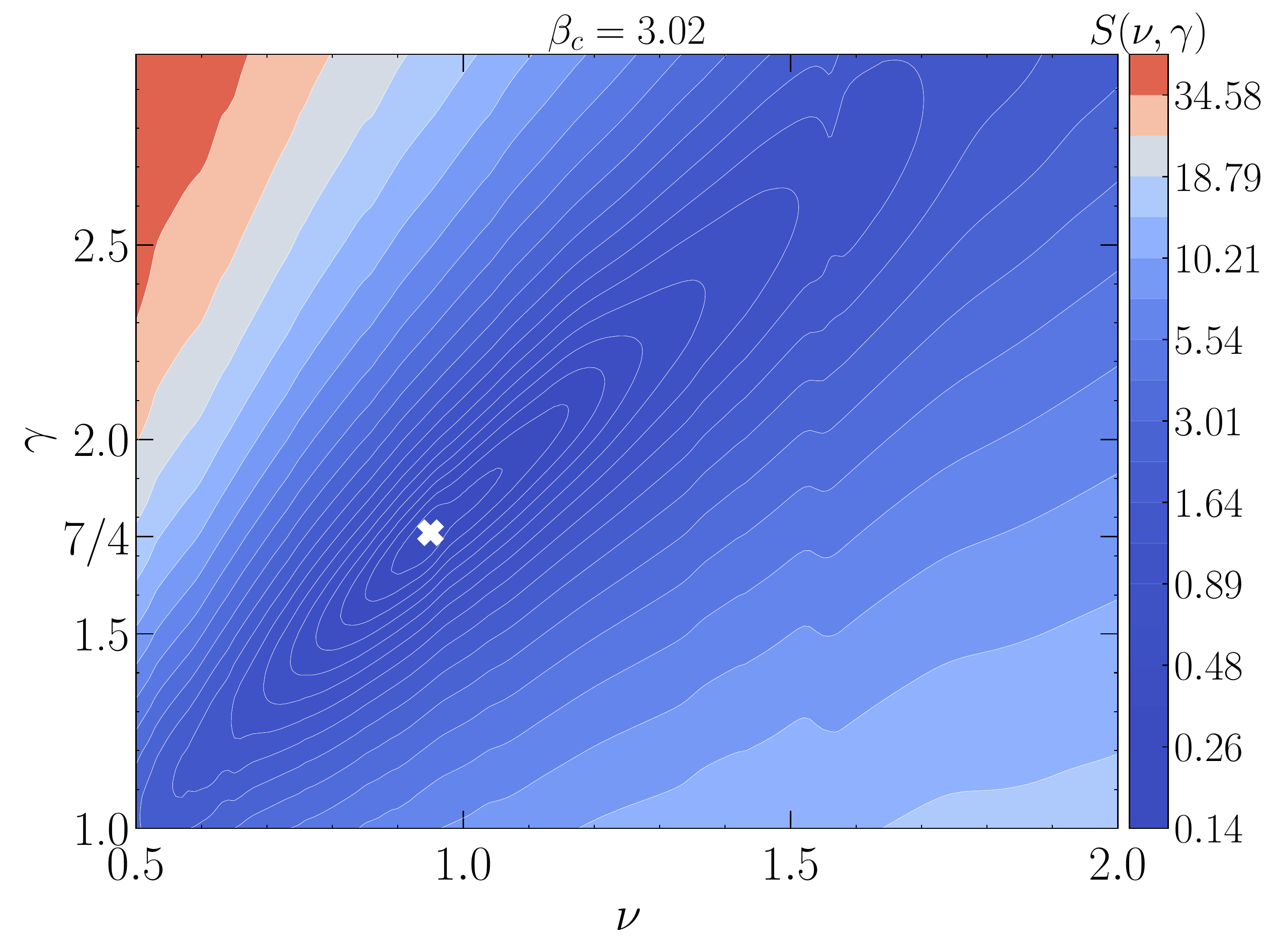} %
\caption{Contour plot of the sum of squared residuals of the least-squares fit $S(\nu,\gamma)$
 of the scaled data for the CDW structure factor $S^{\rm cdw}$ of the Holstein bilayer at
 $t_3=2$ (see Fig.~\ref{fig:bilayerFSS} for the original data). A 16-th order polynomial is used
 to fit the data set, and the critical inverse temperature used is $\beta_c=3.02$. The
 white marker denotes the minimum $S(\nu,\gamma)$  which occurs at $\nu=0.95$ and $\gamma=1.7$, agreeing with the 2D Ising exponents.
 }
\label{fig:contour_plot_gamma_vs_nu} 
\end{figure}

\begin{figure}[t!]
\includegraphics[width=0.99\columnwidth]{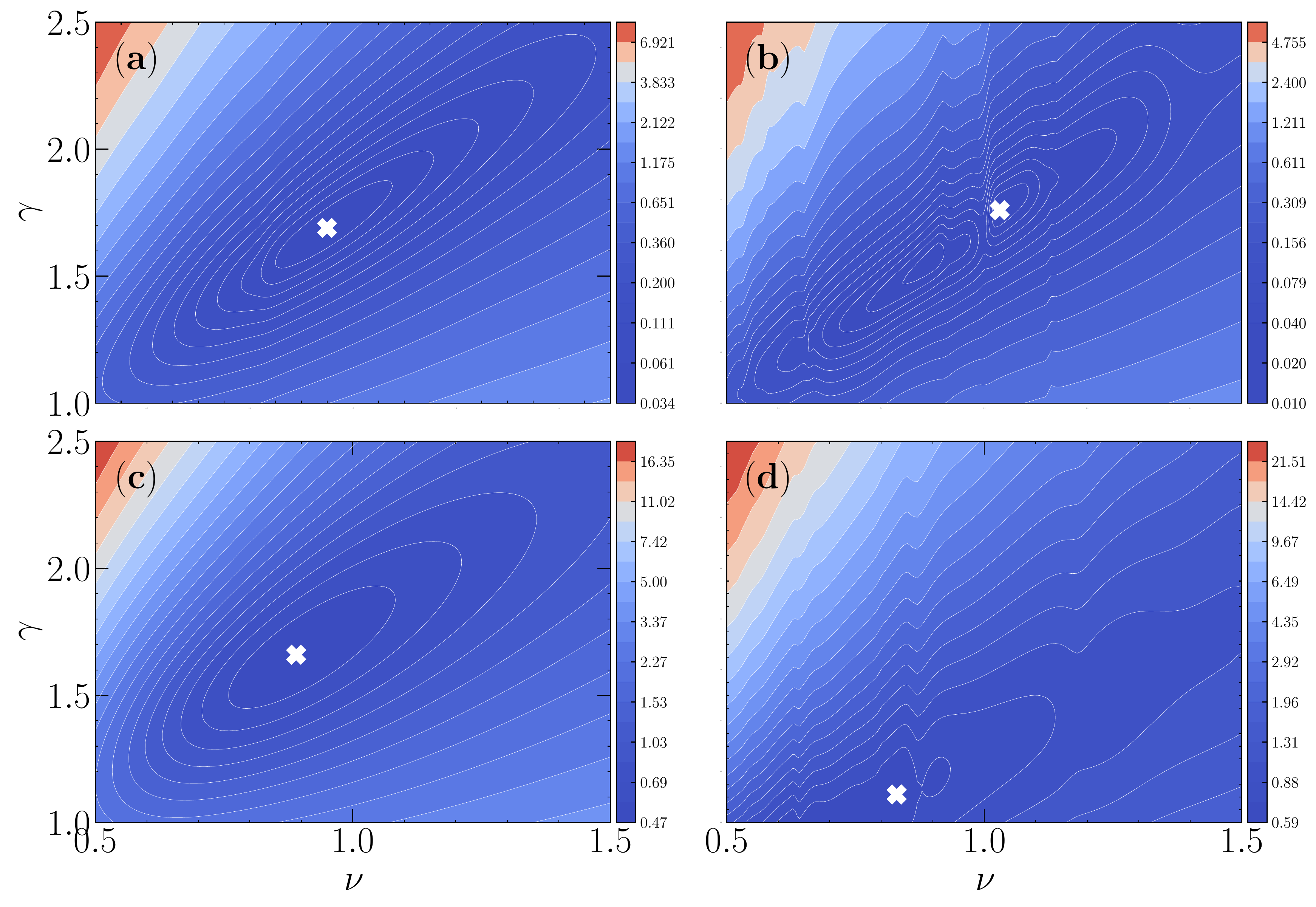} %
\caption{Contour plot of the sum of squared residuals of the least-squares fit $S(\nu,\gamma)$
 of the scaled data for the CDW structure factor $S^{\rm cdw}$ of the Holstein interface,
 top row at $t_3=2$, bottom row at $t_3=1$, left column: layer $l=+1$, right column: layer $l=-1$. (See Fig.~\ref{fig:FSS} for the original data.) A 16-th order polynomial is used
 to fit the data set, and the critical inverse temperature used are listed in Fig.~\ref{fig:fulldatacollapse}. The
 white marker denotes the minimum $S(\nu,\gamma)$ in the displayed range of $\nu$ and $\gamma$. In panels (a,b,c), minimum residuals are at $(\nu=0.95, \gamma=1.69)$, $(\nu=1.03, \gamma=1.76)$ and $(\nu=0.89, \gamma=1.66)$ respectively, whereas in panel (d) the minimum occurs at $\nu=0.83, \gamma=1.11$. 
 Unlike panels (a,b,c),
the $t_3=1$ metallic layer of panel (d) does not fit the known Ising exponents, suggesting
the absence of order in this case.
 }
\label{fig:contour_plot_gamma_vs_nu_1+1} 
\end{figure}

We argue that owing to the symmetry of the order parameter of the CDW phase,
 we make use of the critical exponents pertaining to the 2D Ising universality class in order
 to simplify the FSS of the CDW structure factor $S^{\rm cdw}$. Here, we justify this choice by
 quantitatively extracting the best set of exponents $\nu$ and $\gamma$ that scales the
 curves according to the functional form in Eq.~\ref{eq:FSS}. We start by using the scaled data in a large range of critical exponents; subsequently, for each pair of $(\nu,\gamma)$,
 we proceed with a high-order polynomial fitting of the scaled data, storing the residual
 $S(\nu,\gamma)$ of the fitting procedure. The set of exponents that minimizes $S(\nu,\gamma)$
 is taken as those that characterize the transition. The rationale is that if the dataset 
is well collapsed for a given $(\nu,\gamma)$, a high-order polynomial fit (with
 the number of degrees of freedom much smaller than the number of data points) will
 turn out to have a fairly small error.

Using this procedure, we show in Fig.~\ref{fig:contour_plot_gamma_vs_nu} the contour plot
$S(\nu,\gamma)$ for the data corresponding to the Holstein bilayer at $t_3=2$.
 The minimum residual is indicated by the
 white marker. By observing its variation with slightly different critical inverse
 temperature $\beta_c$, and different polynomial orders used in the fits, we estimate
 $\nu=0.95\pm0.07$ and $\gamma=1.7\pm0.1$, in agreement the 2D Ising exponents 
$\nu_{\rm 2D\ Ising} = 1$ and $\gamma_{\rm 2D\ Ising} = 7/4$. The formation of the CDW phase establishes long range order by breaking the $Z_2$ sublattice symmetry, which is the 2D Ising universality class.

We apply the same analysis to the interface model, as shown in Fig.~\ref{fig:contour_plot_gamma_vs_nu_1+1} (See Fig.~\ref{fig:FSS} and Fig.~\ref{fig:fulldatacollapse} for original data). 
Both layers at $t_3=2$ as well as layer $l=+1$  at $t_3=1$ minimize the residual
close to the 2D Ising critical exponents. However, in the case of layer $l=-1$ at $t_3=1$, the minimum residue is located at $\nu=0.83, \gamma=1.11$ which is far from the 2D Ising exponents. This supports our conclusion that there is no phase transition at $t_3=1$ in layer $l=-1$ since a transition into a phase with long range CDW order necessarily breaks the $Z_2$ symmetry and must be in the Ising universality class.

\section{Induced CDW in Ionic Hubbard Model} \label{sec:analog_model}

We can get additional insight into the Holstein interface by 
considering the following noninteracting, spinless, tight binding Hamiltonian,
\begin{align}
\mathcal{\hat H_{\rm BI-M}} = & -t \sum_{\langle ij
  \rangle,l} \big(
\, \hat c^{\dagger}_{i,l } \hat
c^{\phantom{\dagger}}_{j,l} + {\rm h.c.} \big) + 
\delta \sum_{i} (-1)^i \hat n_{i,1 }
\nonumber \\
 & -t_3 \sum_{j} \big( 
\, \hat c^{\dagger}_{j,1 } \hat c^{\phantom{\dagger}}_{j,-1} + {\rm h.c.} \big) . 
 \label{eq:bim}
\end{align}
Equation
\ref{eq:bim}
describes two bands, labeled by $l=\pm 1$, each with hopping $t$ on a 2D square lattice,
which are hybridized with each other by $t_3$.
Band $l=-1$ is metallic.  At $t_3=0$ it has the 
usual 2D dispersion relation
$\epsilon(k) = -2t \, \big(\, {\rm cos} \, k_x + {\rm cos}\, k_y\, \big)$.
Band $l=+1$ is made insulating by the staggered potential $\delta$,
so that at $t_3=0$ its dispersion relation has two branches,
$E_{\pm}(k) = \pm \sqrt{\epsilon(k)^2 + \delta^2}$.
Both bands of Eq.~\ref{eq:bim} are half-filled (the chemical potential 
$\mu=0$).

\begin{figure}
\includegraphics[width=1\columnwidth]{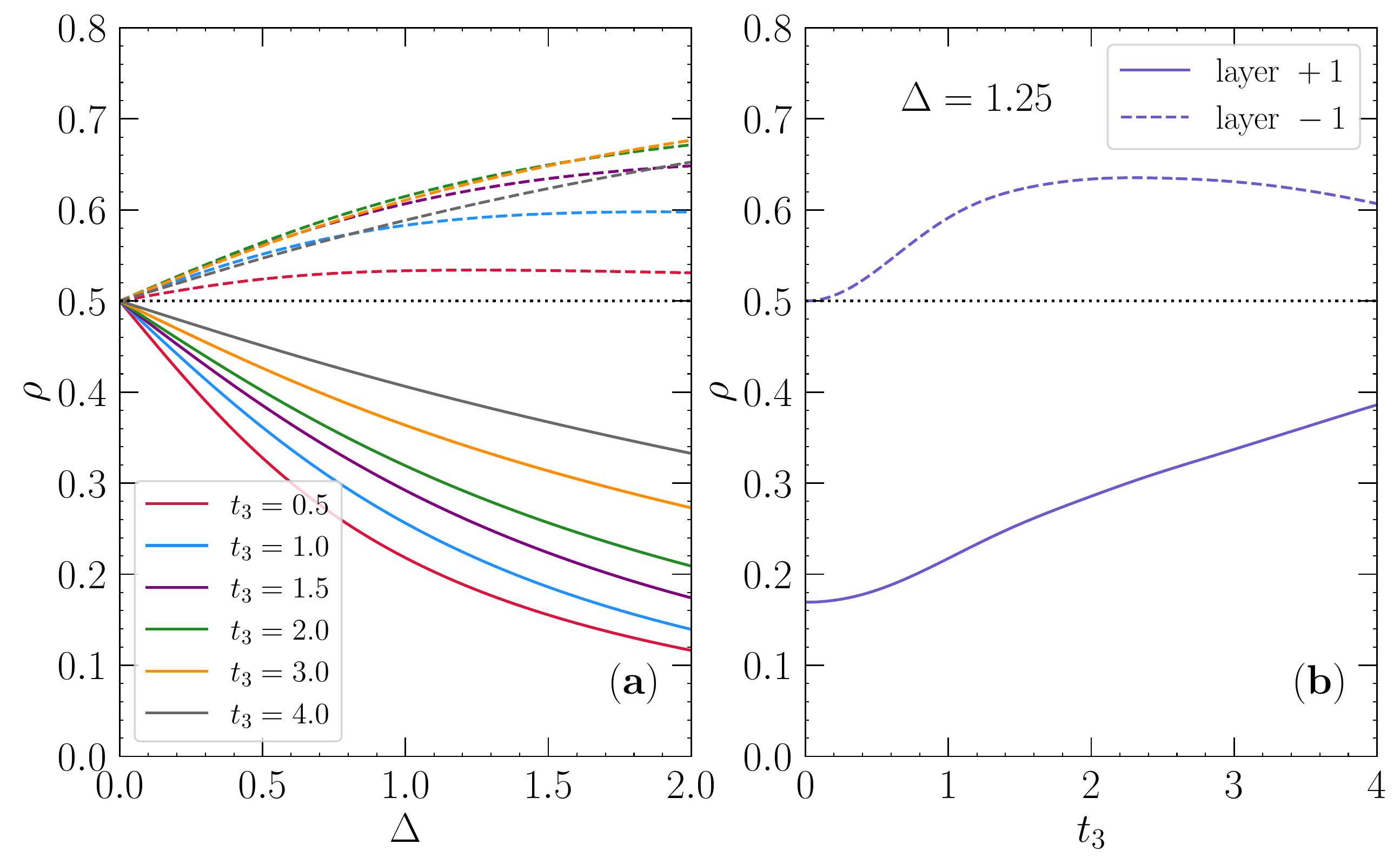} %
\caption{
Solution of the tight binding Hamiltonian, Eq.~\ref{eq:bim}. (a) \underbar{Solid curves} Occupations
 $\rho$ on the $+\delta$ sites of the insulating band as functions of the magnitude of the 
staggered potential $\delta$; \underbar{dashed curves}:  Occupations on the partner sites in
 the metallic band to which those $+\delta$ sites are hybridized by $t_3$.
 (b) \underbar{Solid Curve}:  Occupations on the $+\delta$ sites of the insulating band as
 a function of interlayer hybridization $t_3$. \underbar{Dashed curves}:  Occupations on the partner sites in the 
metallic band to which those $+\delta$ sites are hybridized by $t_3$.
The staggered potential in this case is $\delta=1.25$. In both panels the linear lattice size
 and the inverse temperature  are $L=12$ and $\beta=4$, respectively.}
\label{fig:inducedcdw} 
\end{figure}

In addition to inducing a band gap $2\delta$ in layer $ l=+1$, the staggered potential also
creates a CDW phase, with low occupancy $+\delta$ and high occupancy
$-\delta$ sites. Here, the CDW order is trivial, in the sense of being
induced by an external field, as opposed to arising spontaneously in 
a symmetric Hamiltonian like the Holstein model. Nevertheless we can
still examine how this `artificial' CDW in layer $l=+1$ affects
the site occupations in the metallic band $l=-1$.

Figure~\ref{fig:inducedcdw}(a) gives the occupations on the $+\delta$ sites of
band $l=+1$ as functions of $\delta$ for different $t_3$.
As $\delta$ grows, the occupation of the high energy sites in layer $l=+1$, which 
are directly coupled to the staggered field, get increasingly small (solid curves).
In turn, the occupations of the partner sites on layer $l=-1$ which
are {\it not} coupled to $\delta$ are also shifted from $\rho=\frac{1}{2}$.  
These occupations increase in order to take advantage of the decrease in the Pauli blocking.
What is interesting in the context of the simulations of the Holstein bilayers in the main
part of this paper is that, while the layer $l=+1$ occupations 
steadily return to half-filling with increasing $t_3$, the evolution of the layer $l=-1$
occupations is instead {\it non-monotonic}.  The deviations
of the occupations from half-filling first grow with $t_3$, but then shrink.

This non-monotonicity is seen more clearly in Fig.~\ref{fig:inducedcdw}(b) 
which plots similar occupations as a function of $t_3$ for a fixed $\delta$.
The maximum at intermediate $t_3 \sim 2.28$ is reminiscent of the behavior of
 Fig.~\ref{fig:groundstate}(c), which similarly shows a maximum in the induced CDW order
at intermediate $t_3$ in the metallic layer of the Holstein interface model.
Indeed, the agreement between the values of $t_3$ at which the
induced order is maximal is remarkably {\it quantitative}.
To within error bars, the positions of the maxima are the same,
although the fall-off at large $t_3 $ is more gradual in the BI-Metal interface case.

\bibliography{bibinterface}

\renewcommand{\thefigure}{S\arabic{figure}}
\setcounter{figure}{0}
\renewcommand{\thesection}{S\arabic{section}}
\setcounter{section}{0}
\renewcommand{\theequation}{S\arabic{equation}}
\setcounter{equation}{0}

\end{document}